# Parametric Sequential Method for MRI-based Wall Shear Stress Quantification

*Nina Shokina, Gabriel Teschner, Andreas Bauer, Cameron Tropea, Herbert Egger, Jürgen Hennig, and Axel J. Krafft*

***Abstract*— Wall shear stress (WSS) has been suggested as a potential biomarker in various cardiovascular diseases and it can be estimated from phase-contrast Magnetic Resonance Imaging (PC-MRI) velocity measurements. We present a parametric sequential method for MRI-based WSS quantification consisting of a geometry identification and a subsequent approximation of the velocity field. This work focuses on its validation, investigating well controlled high-resolution *in vitro* measurements of turbulent stationary flows and physiological pulsatile flows in phantoms. Initial tests for *in vivo* 2D PC-MRI data of the ascending aorta of three volunteers demonstrate basic applicability of the method to *in vivo*.**

## I. Introduction

PATHOLOGICAL mechanisms leading to cardiovascular diseases may be discovered by studying blood flow patterns in the cardiovascular system [1]. Various parameters, e.g. volume flow rate, peak velocity, or wall shear stress, are examined with respect to cardiovascular diseases [2]. Wall shear stress (WSS) is the tangential force per unit surface area exerted by flowing blood on the vessel wall [3]. WSS is sensed by endothelial cells located at the blood-vessel wall interface [4] and may serve as a potential biomarker for atherosclerosis, aortic stenosis, aneurysms, bicuspid aortic valves [5]–[7]. See also a review [8] and references therein.

Blood – a shear-thinning non-Newtonian fluid [9] – behaves as an incompressible Newtonian fluid with constant viscosity in larger arteries at high shear rates [10]. In 2D case the WSS at the wall $\partial\Omega$ is defined as

$$\tau_w = -\mu \frac{\partial u}{\partial n}\bigg|_{\partial\Omega},$$

where $\mu$ is the dynamic viscosity, $u$ is the velocity component parallel to $\partial\Omega$, $n$ is the outer normal to $\partial\Omega$ [3]. Phase-contrast Magnetic Resonance Imaging (PC-MRI) [11] is a non-invasive method for measuring blood flow velocity $u$. MRI-based WSS quantification typically requires two steps: 1) the identification of the boundary $\partial\Omega$, 2) the computation of $\partial u/\partial n$ at $\partial\Omega$.

Previous WSS quantification methods locally fitted linear [12]–[14] or parabolic [14] polynomials to the velocity data near the vessel boundary. This approach was generalized in the sectored paraboloid method [15] to the 3D case and using linear interpolation to handle 4D MRI data [16]. These methods relied on sufficiently high spatial resolution and moderate noise near the boundary.

More recently global velocity approximation methods have been developed. An important milestone was the introduction of a method based on B-spline interpolation of the Gaussian filtered measurement data [17]. This method [17] was used in [18]–[20], where the aortic vessel was manually delineated for each 2D analysis plane and WSS estimation was based on a direct interpolation of the local velocity derivative on the segmented vessel contour. In [21] peak-systolic WSS vectors were calculated by fitting the 3D velocity data with B-splines surfaces and computing velocity derivatives on the vessel lumen segmented using the method [17].

In [22] a finite element mesh was introduced, nodal velocity values were derived by cubic interpolation and WSS was computed using the derivative of the finite element function. In [23] the velocity reconstruction using smoothing splines was performed in interior voxels only and the resulting gradient was

This work was supported by the Deutsche Forschungsgemeinschaft (DFG) under grant numbers DFG HE 1875/30-1, DFG TR 194/56-1, and DFG EG 331/1-1.

N. Shokina is with the Department of Radiology, Medical Physics, Medical Center – University of Freiburg, Faculty of Medicine, University of Freiburg, 79106 Freiburg, Germany (e-mail: nina.shokina@uniklinik-freiburg.de).

G. Teschner is with the Institute for Numerical Analysis and Scientific Computing, Department of Mathematics, Technische Universität Darmstadt, 64293 Darmstadt, Germany (e-mail: teschner@mathematik.tu-darmstadt.de).

A. Bauer is with the Institute for Fluid Mechanics and Aerodynamics, Department of Mechanical Engineering, Technische Universität Darmstadt, 64347 Darmstadt-Griesheim, Germany (e-mail: bauer@sla.tu-darmstadt.de).

C. Tropea is with the Institute for Fluid Mechanics and Aerodynamics, Department of Mechanical Engineering, Technische Universität Darmstadt, 64347 Darmstadt-Griesheim, Germany (e-mail: tropea@sla.tu-darmstadt.de).

H. Egger is with the Institute for Numerical Analysis and Scientific Computing, Department of Mathematics, Technische Universität Darmstadt, 64293 Darmstadt, Germany (e-mail: egger@mathematik.tu-darmstadt.de).

J. Hennig is with the Department of Radiology, Medical Physics, Medical Center – University of Freiburg, Faculty of Medicine, University of Freiburg, 79106 Freiburg, Germany (e-mail: juergen.hennig@uniklinik-freiburg.de).

A.J. Krafft was with the Department of Radiology, Medical Physics, Medical Center – University of Freiburg, Faculty of Medicine, University of Freiburg, 79106 Freiburg, Germany. He is now with Siemens Healthcare GmbH, Erlangen, Germany (e-mail: axel.krafft@gmail.com).

extrapolated to the boundary. A similar approach [24] used Sobel filters [25] (a combination of numerical differentiation and orthogonal smoothing) to estimate the velocity gradient in the domain interior, afterwards the extrapolated gradient was evaluated at the boundary. See [8] and references therein on MRI-based WSS quantification methods.

Further research has been dedicated to pre- and post-processing techniques. Pre-processing involves methods for reducing systematic error and background phase distortions in MRI data. Post-processing becomes increasingly important due to an increasing trend to process 4D flow MRI data. For instance, centerline [26] and – for more complicated flow geometries - Laplacian based [27] techniques were developed to define an axial direction and analyze axial and circumferential WSS components.

The assessment of the accuracy of MRI-based WSS estimators is difficult due to the lack of reference data for *in vivo* [2]. WSS values obtained from the 4D flow *in vivo* data with the highest available spatio-temporal resolution might serve as reference [28]. For validation, *in vitro* measurements of controlled flow regimes, Computational Fluid Dynamics (CFD) simulations and Laser Doppler Velocimetry (LDV) experiments [29], [30] can be used. For some flow regimes, analytical solutions or known formulas exist, e.g. the Womersley analytical solution for laminar pulsatile pipe flows [29] and friction factor formula for fully developed turbulent stationary flows in straight pipes [8], [31]. The generic nonlinear regression method for MRI-based WSS quantification [8] automatized the graphical Clauser plot method for fully developed turbulent stationary pipe flows [31], [32].

The accuracy of MR-based WSS estimators depends on the PC-MR acquisition parameters, the boundary identification accuracy, and the accuracy of the velocity reconstruction [2], [14], [23], [28], [33]. PC-MRI-based WSS quantification methods tend to underestimate the WSS values [14], [17], [28], [34]. See [8] and references therein on accuracy and validation of MR-based WSS quantification methods.

The flow velocity reconstruction from MRI data and, subsequent estimation of WSS has three major difficulties.

A first difficulty arises from the flow velocity character. Outside the vessel, the flow velocity and, hence, its derivative are constant zero. A derivative is non-zero inside the flow domain near the boundary, since the shear rate does not vanish. Hence, the derivative features a jump and the velocity profile in the normal direction exhibits a kink at the boundary. This makes an accurate reconstruction of the velocity across the boundary, and in particular its derivatives, difficult by standard methods. E.g., a spline interpolation of the whole velocity function with a zero value for interpolation points outside the vessel, as in [17], exhibits the Gibb's phenomenon [35]: the approximation of the velocity derivative contains a constant error, even for noise-free and infinitely highly resolved data. Therefore, the field of view should be decomposed into a flow region and an exterior domain. The velocity is smooth in the flow domain, in principle allowing an accurate representation using standard approximation methods as it is done in the polynomial fitting methods [12]–[16] as well as in novel approaches [23], [24], [35]. Therefore, a reliable identification of the flow domain from MRI data is crucial.

The second difficulty is the observed dependence on spatial resolution of the underlying velocity data. The limited spatial resolution is the main factor for the variability of MR-derived WSS [2], [14], [17], [23], [28], [33], [36]. The correct near-wall behavior must be appropriately resolved in the data, especially in the case where the velocity profile changes its character in a thin boundary layer. The measurement process incorporates a spatial and temporal local averaging. Most approaches interpret the velocity data as nodal values at voxel centers. However, the voxel value is substantially affected by the flow velocity in a local neighborhood around the voxel center. Furthermore, partial volume effects impact the data in the critical near-wall region. Finally, MRI is not able to capture instantaneous velocity fluctuations or oscillations of the vessel geometry. An accurate modeling of the data acquisition process would minimize the data error and improves stability of the estimates against considerably low data resolution. The temporal resolution may introduce a significant underestimation of WSS in special situations like the emerging of vortex rings in an aneurysm [30]. In the setups investigated in this work temporal resolution has only a minor impact on the WSS estimates [28], [33].

The third difficulty is the evaluation of velocity derivatives at the boundary. The data noise on the boundary is relatively high due to several reasons. The no-slip boundary condition and the continuity of the velocity field lead to low velocities near the boundary. Therefore, the fluid near the boundary is magnetically saturated and its signal-to-noise ratio (SNR) is reduced. Moreover, the velocity encoding (VENC) parameter, that is typically chosen according to the expected maximum velocity to avoid phase aliasing [11], introduces relatively high phase noise near the boundary. The data noise becomes further amplified by numerical differentiation [37]. Therefore, velocity derivatives must be evaluated in a stable manner. In most approaches, regularization by discretization is applied, where the amount of regularization depends on the data resolution only [12]–[16], [24]. This is problematic, since the amount of regularization should be chosen respective to the current flow situation and data fidelity [37] as it was done with the controlled regularization by Gaussian filtering in [17] or smoothing splines in [23].

The proposed method offers an approach to handle all three difficulties appropriately in order to achieve considerable improvements in WSS quantification.

## II. METHODS

**Key ideas**

Our method uses 2D-PC-MRI data given as magnitude and axial velocity values on a regular grid consisting of $N$ voxels to reconstruct both the flow domain $\Omega$, in particular its outer normal $n$, as well as the velocity $u$.

The method has been designed to address the three main

challenges:
- The global velocity is non-smooth, exhibiting a kink at the vessel wall.
- Dependence on data resolution.
- Difficulty in evaluation of velocity derivatives at the boundary.

We use regularization methods [38] to deal with the ill-posedness of the problem indicated by the last two points. To overcome the difficulty with the non-smooth global velocity, we implement a modular sequential approach: first the magnitude data is used to reconstruct the flow geometry, afterwards the identified geometry is used to reconstruct the flow velocity inside the domain. Finally, a WSS estimate is computed. In this work, we briefly present the parametric sequential method with focus on the algorithm. A detailed mathematical analysis of our method, providing error estimates in terms of the measurement error, is given in [40].

**Preprocessing of the magnitude data**

The geometry registration method requires normalized magnitude data $\bar{m}_i = |V_i \cap \Omega|/|V_i|$, which is equal to the fraction of the voxel $V_i$ covered by the flow domain $\Omega$. To extract these values from the original data, we model the magnitude $m_i$ by

$$m_i = \int_{V_i} \rho(x) dx,$$

where $\rho$ is the signal amplitude depending amongst other things on the proton density and the relaxation times. Assuming the signal amplitude in the flow domain $\Omega$ and the exterior domain to be locally nearly constant, we arrive at

$$m_i = \bar{m}_i m_{f,i} + (1 - \bar{m}_i) m_{e,i},$$

with magnitude level $m_{f,i}$ in the fluid and $m_{e,i}$ in the exterior domain around $V_i$. Thus, after selecting the characteristic magnitude levels, the normalized data is obtained as

$$\bar{m}_i = \min\left\{1, \max\left\{0, \frac{m_i - m_{e,i}}{m_{f,i} - m_{e,i}}\right\}\right\}. \quad (1)$$

In the investigated *in vitro* cases, $m_{f,i}$ and $m_{e,i}$ can be chosen independently of $V_i$ by the histogram of the magnitude [39]. For *in vivo* cases, in a first step, a region growing algorithm is applied. For every voxel a surrounding patch is investigated: if all voxels in the patch are picked by the region growing algorithm, the former voxel is designated an interior voxel. If some voxels in the patch are picked, the voxel is called a boundary voxel. Otherwise it is an exterior voxel. For interior and exterior voxels, $\bar{m}_i$ is set to 1 respective 0. For boundary voxels, the normalization (1) is applied with the characteristic magnitude levels $m_{f,i}$ and $m_{e,i}$, selected by the statistics of the nearest interior respective exterior voxel magnitude data. The local magnitude contrast $c_i = m_{f,i} - m_{e,i}$ is used as an indicator for data fidelity.

**Geometry registration**

Given its center of gravity $x_0$ computed from the normalized magnitude $\bar{m}_i$, the flow domain can be parametrically represented using a radius function $R$ as shown in Fig. 1a by

$$\Omega(R) = \{x \in \mathbb{R}^2 : |x - x_0| < R(\varphi(x - x_0))\},$$

where $\varphi(x)$ is the angle between first axis and the vector $x$. To determine the radius function $R$, such that $\Omega(R)$ is the flow domain, we introduce the virtual measurement operator

$$F_{geo}(R)_i = |V_i \cap \Omega(R)|/|V_i|$$

and the Tikhonov functional

$$J_{geo}(R) := \frac{1}{N} \sum_{i=1}^{N} \bar{c}_i |F_{geo}(R)_i - \bar{m}_i|^2 + \alpha \|R''\|^2_{L^2(0,2\pi)}. \quad (2)$$

The first term forces accordance to the data, where the data fidelity indicator $\bar{c}_i = c_i/\max(c_i)$ is the normalized magnitude contrast $c_i = m_{f,i} - m_{e,i}$ ($\bar{c}_i = 1$ for exterior and boundary voxels and *in vitro* cases). The second term ensures the smoothness of $\Omega(R)$. $R$ is computed by minimizing $J_{geo}$ over the space of functions of type

$$R(\varphi) = b_0 + \sum_{k=1}^{n} a_k \sin(k\varphi) + b_k \cos(k\varphi).$$

In practice, the forward operator has to be smoothed to enable gradient methods for the minimization. The regularization parameter $\alpha$ (2) to be chosen carefully (here $\alpha = 0.1$ was selected) and $n$ was chosen to be sufficiently large avoiding discretization errors. For details we refer to [39].

For a non-circular geometry, the registration is illustrated in Fig. 1b. Table I lists the relative geometry errors for a synthetic data resolution series with the fixed SNR=10.

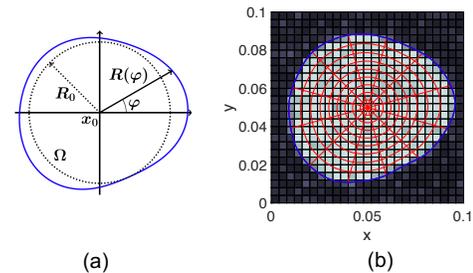

(a) (b)
Fig. 1. (a) the parametric representation of geometry, (b) magnitude image of the synthetic data with illustration of the identified geometry (blue) and the domain transformation (red).

TABLE I. RELATIVE ERRORS OF GEOMETRY REGISTRATION FOR THE SYNTHETIC DATA RESOLUTION SERIES

| voxels in *x*-direct. | 200 | 100 | 50 | 25 |
|---|---|---|---|---|
| relative error | $1.43 \times 10^{-2}$ | $4.36 \times 10^{-2}$ | $9.93 \times 10^{-2}$ | $3.09 \times 10^{-1}$ |

**Velocity approximation**

Contrary to the global velocity field, the velocity field restricted to the flow domain $\Omega$ is smooth and can be approximated by standard techniques. The parametrization $R$ and $x_0$ of the identified geometry defines a diffeomorphism $\phi$, mapping the unit disk $B = \{x \in \mathbb{R}^2 : \|x\| < 1\}$ to the flow domain $\Omega$. With $R_0 = \min_{0 \leq \varphi \leq 2\pi} R(\varphi)$ the transformation $\phi$ is given by

$$\phi : \begin{pmatrix} r\cos(\varphi) \\ r\sin(\varphi) \end{pmatrix} \mapsto x_0 + (R_0 r + (R(\varphi) - R_0)r^4)\begin{pmatrix} \cos(\varphi) \\ \sin(\varphi) \end{pmatrix}.$$

As illustrated in Fig. 1b, the deviations $R(\varphi) - R_0$ from the minimal radius become apparent in the boundary region only, leading to a smooth transformation everywhere.

Every arbitrary function $v : B \to \mathbb{R}^n$ defined on the unit disk $B$ can be associated with a function $u : \Omega \to \mathbb{R}^n$ on the flow domain by $u(x) = v(\phi^{-1}(x))$. Modelling the velocity measurement by the averaging

$$u_i = \frac{1}{|V_i \cap \Omega|} \int_{V_i \cap \Omega} u(x) dx$$

leads to the forward operator

$$F_{flow}(v)_i = \frac{1}{|V_i \cap \Omega|} \int_{V_i \cap \Omega} v(\phi^{-1}(x)) dx.$$

The associated Tikhonov functional reads

$$J_u(v) = \frac{1}{|\Omega|} \sum_{|V_i \cap \Omega| > 0} |V_i \cap \Omega| |F_{flow}(v)_i - u_i|^2 + \beta \|\Delta v\|^2_{L^2(B)}, \quad (3)$$

$v$, and therefore, the velocity approximation $u = v \circ \phi^{-1}$, is obtained by minimizing the Tikhonov functional over the truncated spectral space

$$V_M = \text{span}\left\{ v \in L^2(B) \middle| \begin{cases} -\Delta v = \lambda v & \text{in } B \\ v = 0 & \text{on } \partial B \end{cases}, \lambda \leq M \right\}. \quad (4)$$

Note, that the approximation incorporates the no-slip condition. A typical velocity exhibiting a boundary layer shows high derivatives in normal direction and small derivatives in the angular direction (Fig. 2). The former requires a low regularization parameter (here $\beta = 10^{-8}$). A larger smoothing in the angular direction is applied by discretization: using polar coordinates the basis functions in (4) are of the type $v(r,\varphi) = \Re(e^{k\varphi i})f(r)$ [40]. The basis function $v$ is called low-oscillatory for $k \leq N_\varphi$ and high-oscillatory for $k > N_\varphi$. The splitting number is related to the angular size $\Delta\varphi$ of the smallest resolved structure by $N_\varphi = 2\pi/\Delta\varphi$. We use in vitro $N_\varphi = 0$ and in vivo $N_\varphi = 4$ ($\Delta\varphi = 45^0$). To not spoil the solution with discretization errors, we include low-oscillatory basis functions with eigenvalue $\lambda \leq M_1 = 200$. Only in vitro, we additionally include high-oscillatory basis functions with $\lambda \leq M_2 = 20$ to compensate for deviations from the symmetric velocity profile.

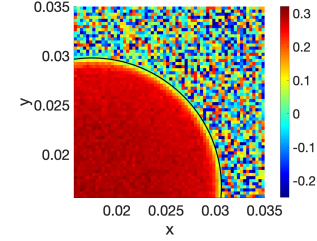

Fig. 2. A visible boundary layer in the velocity image of the laminar nonstationary flow with identified geometry.

**Evaluation of wall shear stress**

The unit outer normal $n$ and the gradient $\nabla u$ of the velocity field are analytically computable at any point from the known geometry parametrization $x_0$ and $R$, the domain transformation $\phi^{-1}$ and the velocity approximation $v$ respective their derivatives in a stable manner [39]. At the boundary point $x$ with angle $\varphi = \varphi(x - x_0)$ the WSS estimate is given by $\tau_w(\varphi) = -\mu n(\varphi)^T (\nabla u)(x)$.

For nonstationary flows we also estimate the oscillatory shear index (OSI) [41] – a spatially distributed quantity measuring oscillations of the WSS over time at the boundary parametrized by the angle $\varphi$ in our approach:

$$\text{OSI}(\varphi) = \frac{1}{2}\left(1 - \frac{\left|\int_0^T \tau_w(\varphi,t)dt\right|}{\int_0^T |\tau_w(\varphi,t)|dt}\right). \quad (5)$$

The absolute physical position of boundary may slightly move over time due to temporal variations of the geometry. Note, that only measurements of the axial velocity were conducted and, therefore, incorporated into the OSI computation, which may lead to a distorted OSI especially *in vivo*, were the circumferential component of the WSS is not negligible.

**Validation *in vitro***

*In vitro* flow setups were turbulent stationary flows in a glass pipe (∅ 25.9 mm) and laminar physiological pulsatile flows (mimicking flow in the human aorta [29]) in an acrylic pipe (∅ 26 mm). A pipe was placed inside a 3T whole-body scanner (MAGNETOM Prisma, Siemens Healthineers, Erlangen, Germany) along the magnet's center line. Flow regimes were provided by two pumps (RMMSI, Sondermann, Köln, Germany, CardioFlow MR 5000, Shelley Med., Canada), located outside the MR scanner room and connected via plastic hoses and straight in-flow pipes (length ≈ 2 m) to ensure fully developed flow conditions (Fig. 3). A fluid was pure water at room temperature doped with copper sulfate (1g/l) [42]. The setups were carefully controlled to ensure the reproducibility of the experiments. More details are given in [29]. A PC-MRI sequence was based on a spoiled 2D gradient echo sequence with bipolar velocity encoding along the slice selection direction. 2D PC-MR images were acquired in a plane oriented perpendicular to the pipe axis for different in-plane resolutions with velocity encoding along the through-plane direction.

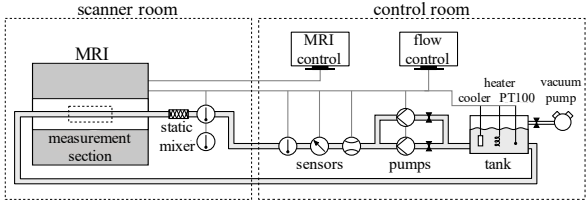

Fig. 3: Experimental setup for MRI *in vitro* measurements [29].

The reference values for the turbulent flows were computed by the friction factor formula [8], [31] and the generic nonlinear regression method [31], for the laminar flows – from the LDV experiments [29] and the Womersley analytical solution [29]. For comparison, the WSS was obtained by the methods of Potters et al. [23] and Stalder et al. [17] (called as *flow-tool*) using our geometry identification. Both methods are widely used, serve as kernel for many other WSS estimators [8], and can be considered as the current standard.

## III. RESULTS

**Turbulent stationary flows**

Turbulent stationary flows in a straight pipe were provided at Reynolds numbers $Re = 8060, 5370, 3000$. VENC = 0.25 m/s was chosen for all $Re$ in order to better resolve the velocities close to the vessel boundary. The images were acquired for nine different in-plane resolutions (range: 0.30 mm x 0.30 mm – 1.50 mm x 1.50 mm, relative resolution 17,3-86,3 voxels). Other scan parameters were TR/TE = 17.8-18.6/5.66-5.68 ms, FOV was 96 mm × 96 mm, slice thickness was 3 mm, flip angle 5°.

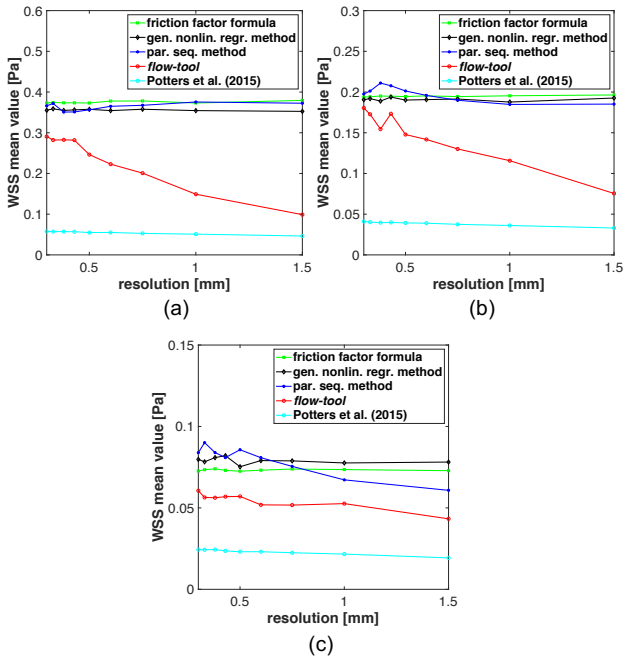

Fig. 4. MR-derived WSS estimates by the parametric sequential method, *flow-tool*, the method [23], the friction factor formula [8], [31] and the generic nonlinear regression method [31] for turbulent stationary flows with (a) $Re = 8060$, (b) $Re = 5370$, (c) $Re = 3000$ and different in-plane resolutions.

The geometry was identified accurately (relative approximation errors of 0.01-0.08% for turbulent stationary flow with $Re = 5370$), i.e. sub-voxel resolution is achieved. As observed from Fig. 4, in contrast to *flow-tool* the estimates of the parametric sequential method deviate only slightly from the reference values and reveal no dependence on the data resolution. The relative deviation between the values obtained by the parametric sequential method and the friction factor formula was 4.2%±2.5%. The deviation between the two reference values was 2.1%±1.0%, 26.3%±17.3% between *flow-tool* and the friction factor formula, 80.3%±1.4% between the method of Potters et al. [23] and the friction factor formula. The method [23] showed little dependence on the resolution, but large deviation from the reference data. This method was calibrated using a stationary Poiseuille flow only, hence, being highly accurate for these flow regimes, but revealing severe underestimations in the case of boundary layers.

Table II shows the influence of SNR on the WSS estimates using variously averaged data of the turbulent stationary flow.

TABLE II. INFLUENCE OF SNR: MEAN ERROR AND STANDARD DEVIATION OF THE WALL SHEAR STRESS VALUES (TURBULENT STATIONARY FLOW, $Re = 5370$, THE IN-PLANE RESOLUTION 0.3 MM X 0.3 MM) FOR VARIOUSLY AVERAGED DATA; REFERENCE VALUES WERE OBTAINED BY THE FRICTION FACTOR FORMULA FOR THE AVERAGED 12 MEASUREMENTS

|  | 12 single meas. | 20 random twos of meas. | 20 random fours of meas. | 20 random eights of meas. |
|---|---|---|---|---|
| mean error | $2.232234 \cdot 10^{-1}$ | $1.976384 \cdot 10^{-1}$ | $2.154909 \cdot 10^{-1}$ | $1.912649 \cdot 10^{-1}$ |
| std. dev. | $1.047512 \cdot 10^{-1}$ | $9.357256 \cdot 10^{-2}$ | $5.633551 \cdot 10^{-2}$ | $2.213763 \cdot 10^{-2}$ |

**Laminar nonstationary (physiological pulsatile) flows**

Two physiological pulsatile flows in a straight pipe with maximal Reynolds numbers $Re_{\max} = 3952$ (resting state) and $Re_{\max} = 7651$ (exercising state) mimicked realistic flow conditions of the human aorta [29]. A Womersley number $Wo = 20.3$ was taken as characteristic for the ascending aorta [43]. The images were acquired for four different in-plane resolutions (range: 0.30 mm x 0.30 mm – 1.56 mm x 1.56 mm, relative resolution 16,7-86,7 voxels). Other scan parameters for resting state were VENC = 0.15 m/s, TR/TE = 70.4-72.0/5.7795-5.68 ms, FOV was 32 mm × 32 mm – 96 mm × 96 mm, slice thickness was 3 mm, flip angle was 10°. The scan parameters for exercising state were VENC = 0.25 m/s, TR/TE = 64.4-68.0/5.24-5.25 ms, FOV was 32 mm × 32 mm – 160 mm × 160 mm, slice thickness was 3 mm, flip angle was 10°.

Our method shows low dependence on the data resolution in contrast to *flow-tool* and good agreement with the reference data in contrast to *flow-tool* and the method [23] (Fig. 5, 6). However, the method reveals a phase difference to the reference values during the phase around the second positive WSS peak. One possible reason is that the inertia driven velocities in the middle of the pipe influence the WSS estimate. These velocities lag behind the velocities in the boundary layer, driven by viscous forces and follow the changing pressure gradient faster. Nevertheless, this phase difference does not seriously affect important quantities such as peak WSS and OSI, which is in a good agreement with the reference (see Table III).

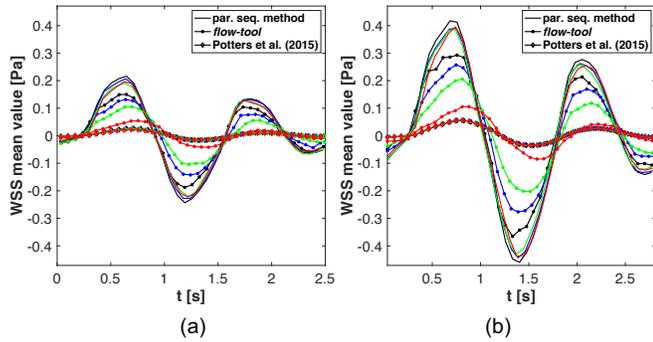

Fig. 5. MR-derived WSS estimates by the parametric sequential method (no markers), *flow-tool* (circles) and the method [23] (diamonds) for the laminar physiological pulsatile flows with (a) $Re_{\max} = 3952$, (b) $Re_{\max} = 7651$; resolutions: black lines – 0.30 mm x 0.30 mm, blue lines – 0.52 mm x 0.52 mm, green lines – 0.78 mm x 0.78 mm, red lines – 1.56 mm x 1.56 mm.

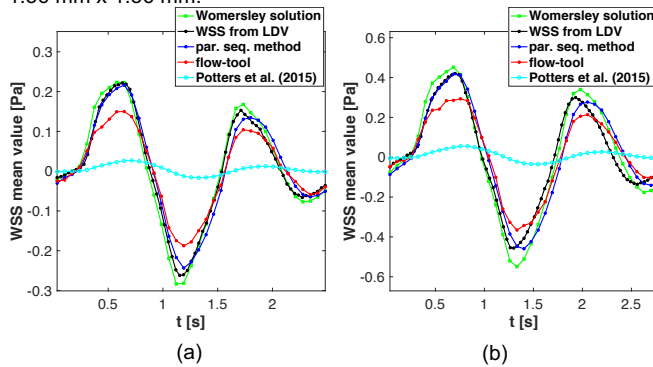

Fig. 6. MR-derived WSS estimates by the parametric sequential method (no markers), *flow-tool* (circles) and the method [23] (diamonds) for the in-plane resolution 0.3 mm x 0.3 mm, the WSS values obtained from LDV [29], the Womersley analytical solution [29] for the laminar physiological pulsatile flows with a) $Re_{\max} = 3952$ and b) $Re_{\max} = 7651$.

TABLE III. OSCILLATORY SHEAR INDEX (OSI) FOR PHYSIOLOGICAL PULSATILE FLOWS

| in-plane resolution | $Re_{\max} = 3952$ | $Re_{\max} = 7651$ |
|---|---|---|
| 0.30 mm × 0.30 mm | 0.4614 | 0.4580 |
| 0.52 mm × 0.52 mm | 0.4535 | 0.4686 |
| 0.78 mm × 0.78 mm | 0.4655 | 0.4649 |
| 1.56 mm × 1.56 mm | 0.4638 | 0.4636 |
| Womersley solution | 0.4737 | 0.4728 |

**Influence of regularization parameters**

As expected, the method is sensitive to the choice of the regularization parameter $\beta$ (Fig. 7a). $\beta = 10^{-8}$ yields convincing results for physiological flows (Fig. 5-6). Assuming SNR to be independent of the Reynolds number, this is expected, since the velocity profiles were found to be close to the Womersley solution [29] and thus linearly depending on the Reynolds number, too. Also, the estimates show only little dependence on the resolution, which is in agreement with the observation of a constant optimal regularization parameter in the preliminary study [39]. Thus, $\beta = 10^{-8}$ is an appropriate choice for *in vivo* studies of the axial aortic flow, independently of the Reynolds number and the MRI resolution. Discretization errors become important for far lower discretization parameter $M_1 < 100$ only (Fig 7b).

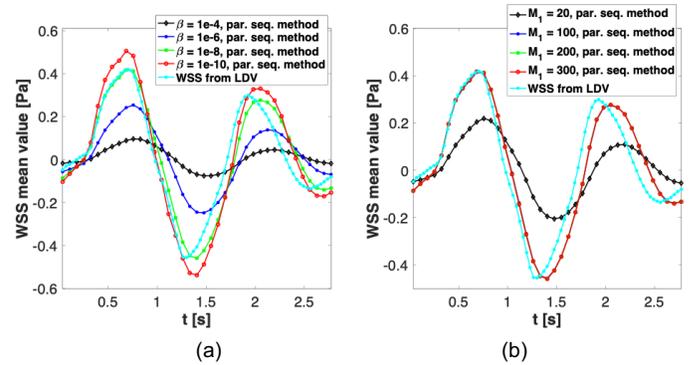

Fig. 7. MR-derived WSS estimates by the parametric sequential method for physiological pulsatile flow with $Re_{\max} = 7651$, in-plane resolution 0.30 mm x 0.30 mm and different values of (a) Tikhonov regularization parameter $\beta$, (b) discretization parameter $M_1$.

**Application *in vivo***

The parametric sequential method was applied to *in vivo* 2D PC-MRI data of the ascending aorta acquired for three healthy volunteers (men, 42, 29 and 20 years old) using ECG-triggering and breathing navigator gating. A scan plane was placed axially in the ascending aorta (Fig. 8) and the images were acquired for four different in-plane resolutions (range: 0.6 mm x 0.6 mm – 1.1 mm x 1.1 mm). The VENC was 1.5 m/s, TR/TE were 21.6-22.8/2.9 ms, FOV was 500 mm × 250 mm, slice thickness was 5 mm, flip angle was 15°.

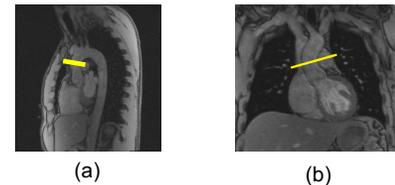

Fig. 8. Scan plane: (a) sagittal plane, (b) coronal plane.

*In vivo*, the vessel boundary is well identified for various image contrasts and SNR during the cardiac cycle (Fig. 9). A healthy aortic vessel wall is 1-2 mm thick, i.e. for the resolution 0.6 mm x 0.6 mm (Fig. 9) the vessel wall covers about 2-4 voxels. The wall blurring might occur due to the aorta motion and partial volume effects. For correct WSS assessment the registered boundary has be maximally close to the inner side of the wall. Table IV shows that if the geometry in Fig. 9b is enlarged by 1-2 voxels, then the WSS estimate increases and even changes sign which is unphysical.

The WSS estimates of the parametric sequential method are substantially larger than the values provided by *flow-tool* and the method [23] (Fig. 10) and in accordance with the order of magnitudes reported in the literature: 3.106 ± 0.479/3.385 ± 1.553 Pa [44] and 2.23±1.04 Pa [36]. In contrast to *flow-tool* our method shows little dependence on the spatial resolution.

TABLE IV. ASCENDING AORTA CROSS-SECTION AREA AND WALL SHEAR STRESS VALUES FOR GEOMETRY IDENTIFIED BY THE PARAMETRIC SEQUENTIAL METHOD AND ITS ENLARGEMENTS BY 1 AND 2 VOXELS

| | original geometry | +1 voxel | +2 voxels |
|---|---|---|---|
| vessel cross-section area | $0.556 \cdot 10^{-3}$ m$^2$ | $0.608 \cdot 10^{-3}$ m$^2$ | $0.662 \cdot 10^{-3}$ m$^2$ |
| wall shear stress | -1.04 Pa | -0.45 Pa | 0.06 Pa |

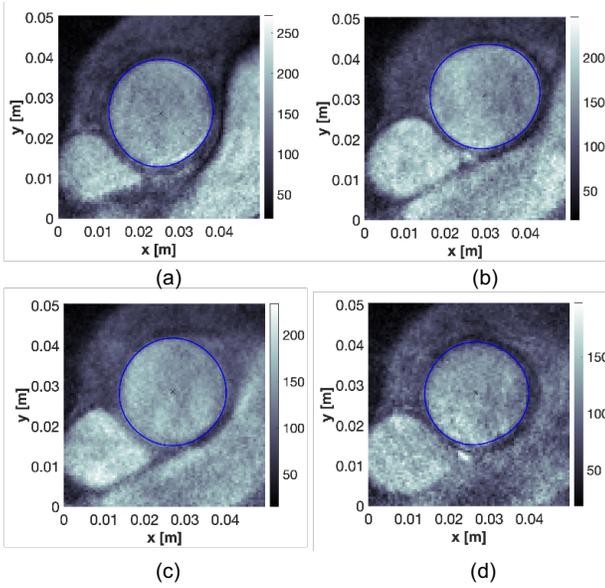

Fig. 9. Magnitude images of the ascending aorta with the vessel boundary identified by the parametric sequential method for: (a) time step 6, (b) time step 17, (c) time step 23, (d) time step 40.

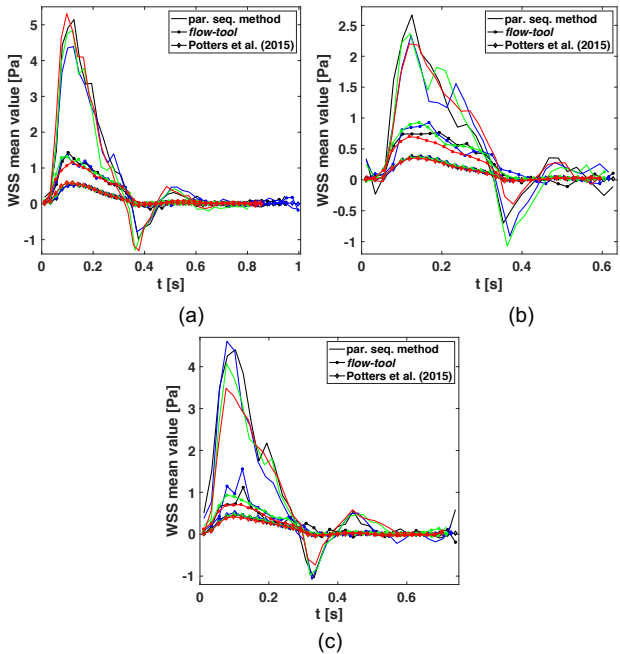

Fig. 10. MR-derived WSS estimates by the parametric sequential method (no markers), *flow-tool* (circles) and the method [23] (diamonds) for *in vivo* MRI measurements for three volunteers; resolutions: black lines – 0.30 mm x 0.30 mm, blue lines – 0.52 mm x 0.52 mm, green lines – 0.78 mm x 0.78 mm, red lines – 1.56 mm x 1.56 mm.

## IV. Discussion

In this work, a new parametric sequential method for MRI-based wall shear stress quantification was presented. The method was validated on well controlled *in vitro* measurements and applied to *in vivo* data of the ascending aorta of three volunteers.

Due to the presence of boundary layers, both investigated *in vitro* setups are challenging for generic WSS estimators and relevant for the *in vivo* situation. The parametric sequential method copes well with both *in vitro* cases. The resulting WSS estimates are robust against low data resolution and show low deviation from the reference values. For the physiological flows, the important quantities peak WSS and OSI are estimated especially well. Although the results reveal high sensitivity on the regularization parameter, the good performance of the parametric sequential method with fixed regularization parameter $\beta = 10^{-8}$ in all setups validates it as a promising tool to estimate WSS *in vivo*. The parametric sequential method successfully identifies the inner side of the vessel wall needed for correct WSS assessment. Based on comparison to the literature, the method in fact provides reliable estimates of WSS in the ascending aorta.

During our investigations the accurate identification of the flow domain turned out to be crucial for reasonable WSS estimates, as it is expected [39]. This does not affect the estimates *in vitro*, due to the highly accurate geometry identification, but requires careful adjustments of the region growing parameters of the magnitude pre-processing *in vivo*. Furthermore, the accuracy of the velocity reconstruction may be further enhanced by including fluid dynamical models in the smoothing process [45]. This requires the extension of our method to 3D, that is possible as explained in [39], but requires some modifications to establish computational feasibility.

### Acknowledgment

Nina Shokina thanks Dr. Maximilian Russe and Thaís Araujo Pérez (Department of Radiology, Center for Diagnostic and Therapeutic Radiology, Medical Center – University of Freiburg, Faculty of Medicine, University of Freiburg, Freiburg, Germany) for the help with the quality assessment of the *in vivo* geometry identification and Adriana Komanscek Department of Radiology, Medical Physics, Medical Center – University of Freiburg, Faculty of Medicine, University of Freiburg) for the help with the *in vivo* data acquisition.